\documentclass[doublecol]{epl2}
\newcommand{\figwidth}{3.275 in}
\newcommand{\figwidthb}{2.725 in}
\title{\bf Possible role of $^3He$ impurities in solid $^4He$}

\author{Efstratios Manousakis}

\institute{                    
  \inst{1} Department of Physics and MARTECH, Florida State University, 
Tallahassee, FL 32306-4350, USA and\\
  \inst{2} Department of  Physics, University of Athens, 
Panepistimiopolis, Zografos, 157 84 Athens, Greece.

}

\abstract{
We use a  quantum lattice  gas model  to describe  essential
aspects  of the motion of $^4He$ atoms and of $^3He$  impurities in  
solid $^4He$. This study suggests that $^3He$ 
impurities  bind to defects and promote $^4He$ atoms to
interstitial sites which can turn the bosonic quantum disordered crystal
into a metastable supersolid. It  is suggested that defects
and interstitial atoms are produced during the solid  $^4He$ 
nucleation process where the role of $^3He$ impurities (in addition to
the cooling rate) is known to be
important even at very small (1 ppm) impurity concentration. 
It is also proposed that such defects can form a glass phase
during the $^4He$ solid growth by rapid cooling.}
\begin{document}

\maketitle
%\section{Introduction}
 Kim and Chan\cite{chan} (KC) using the torsional oscillator 
technique found a decrease in the resonant period of solid $^4He$ confined 
in porous vycor glass and in bulk solid helium
below $200 mK$, indicating the possible onset 
of superfluidity in solid helium. 
The experimental results of KC have been independently confirmed by
Rittner and Reppy\cite{reppy} (RR) using a different geometry to confine the
solid. In addition, RR observed
that the superfluidity of solid helium can be significantly influenced
or eliminated by annealing of the solid helium sample.
Because of these history-dependent results and the negative
results in attempts to drive flow by pressure\cite{day}, 
the interpretation of the results of KC 
%as a superflow of a superfluid component in an 
%ideal quantum crystal, 
is subject to debate.

Furthermore, it has been observed that the superfluid
response and superfluid fraction are strongly dependent 
on the amount of isotopic $^3He$ impurities which
exist in the naturally available helium.
When, for example, the naturally occurring concentration of
$^3He$ impurities of about one part per million is reduced
to less than one part per billion\cite{chan4} the superfluid fraction is
reduced from about $1\%$ to approximately $0.03 \% $. When 
the $^3He$ impurity concentration is increased to about $0.1\%$ the
superfluid fraction vanishes\cite{chan}.
Previously, Ho {\it et al.}\cite{goodkind} found an anomalous behavior
of the acoustic attenuation and velocity in solid $^4He$ below 
$\sim 200 mK$ at a low concentration of $^3He$  
impurities.

The possibility of superfluidity of solid $^4He$ 
has been extensively discussed\cite{andreev,leggett,prokofev,ceperley}.
It~is~of~fundamental value for our understanding of this and
related phenomena because of the implication that there 
can be coexistence between spatial 
and momentum space order\cite{penrose}. 

% Here you need to explain what you will do and what you find.

In this paper we  study the role of very low density of impurities  
in a bosonic hard core solid. While the theoretical studies of the
role of impurities in solid helium has a long 
history\cite{andreev,kagan},
the present study, based on an analogy with models of doped quantum
antiferromagnets and by using rather recently developed techniques
for such systems, sheds some ``new'' light on the
problem.  Our model, that describes the impurity motion in an 
otherwise ideal quantum bosonic crystal,
maps to a quantum spin model with antiferromagnetic (AF) coupling and
 AF order in one direction and ferromagnetic coupling in the perpendicular
direction with impurities moving through the lattice. 
The impurity motion between sub-lattices couples to quantum 
fluctuations of these pseudo-spin degrees of freedom 
which correspond to the boson hopping from an
occupied site of the solid to an empty interstitial site.
We find that, for the limit which describes the case of
solid $^4He$, interstitial atoms are well-defined delocalized
excitations. It is suggested that during the $^4He$ solid nucleation process,
the $^3He$ impurities stabilize point defects such as dislocations or 
disclinations  at the solid-to-liquid
interface which become lines of defects as the solid grows.
These defects can become highly entangled and can form a 
glass phase at low temperature. The $^3He$ impurities become
bound to these defects  and promote mobile interstitial
$^4He$ atoms into the disordered solid. We find that such interstitial
atoms can condense and may be responsible for the observed
behavior\cite{chan}.

Let us consider a lattice gas model to describe the bosonic solid and the added
impurities. In such a  model,  we need to consider the interstitial 
sites as part of the lattice and, thus, the ideal quantum solid 
containing no vacancies and no impurities, corresponds to a 
fractionally occupied lattice. 
For example, the ideal triangular solid corresponds to the 
case of 1/3 filling, namely, to a $\sqrt{3}\times \sqrt{3}$ ordered solid
and the ideal square lattice solid corresponds to the  
$\sqrt{2}\times\sqrt{2}$ 
checkerboard solid, i.e., 1/2 filling of the lattice with bosons. 
Our model Hamiltonian describing a bosonic quantum solid (such as solid
$^4He$)  and a small concentration of impurities 
(such as  $^3He$ atoms) may be written as follows:
\begin{eqnarray}
 \hat {\cal H}=\hat {\cal H}_B + \hat {\cal H}^* + \hat {\cal H}_{int}
-\mu \sum_i \hat N_i - \mu^* \sum_i \hat n_i, \label{model}\\ 
%\end{eqnarray}
%\begin{eqnarray}
\hat {\cal H}_B=\sum_{<ij>} [-t_b (B^{\dagger}_i B_j + h.c.)
+ V \hat N_i \hat N_j ]  +  U \sum_i  \hat N_i^2 ,
\label{bosonic} \\
\hat {\cal H}^*=\sum_{<ij>}[-t (c^{\dagger}_i c_j + h.c.) +  
W\hat n_i \hat n_j] + U^* \sum_i \hat n^{2}_i,\label{impurity} \\
\hat {\cal H}_{int}=V^* \sum_{<ij>}  (\hat N_i \hat n_j + \hat n_i \hat N_j)
+ U' \sum_i \hat n_i \hat N_i, \label{interaction}
\end{eqnarray}
${\cal H}_B$ is the Hamiltonian of the bosonic solid, and $B^{\dagger}_i(B_i)$ 
are boson creation (annihilation) operators and
$\hat N_i= B^{\dagger}_i B_i$. 
The fermion operators $c^{\dagger}_i (c_i)$  create (or annihilate) 
impurities on the site $i$ and we have suppressed their spin degree
of freedom for simplicity.  Here, $n_i$ is
the impurity number operator.
We will consider the $U\to \infty$, $U^* \to \infty$ and $U' \to \infty$
 limits (single-site occupation subspace) because of the the 
hard-core interaction. $V,V^*$ and $W$ are positive 
because of the additional energy cost to 
place an atom in an interstitial site. 
%For the case of $^3He$ impurities in solid $^4He$, $V < V^*$.

%\section{The effective Hamiltonian}

%\subsection{Pure bosonic solid}
In the absence of impurities the well-known pure bosonic 
Hamiltonian (\ref{bosonic}), 
in the limit of $U \to \infty$ and under the transformation
$B^{\dagger}_i \to S^+_i, B_i \to S^-_i, N_i \to S^z_i - 
{1 \over 2}$,
where $(\hat S^x_i,S^y_i,S^z_i)$ are spin-1/2 operators,
reduces to the anisotropic spin-1/2 Heisenberg model\cite{RMP}
\begin{eqnarray}
\hat {\cal H}_B = \sum_{<ij>} [ J S^z_i S^z_j + 
J_{xy} (S^x_i S^x_j + S^y_i S^y_j)]
- H \sum_i S^z_i,
\label{heisenberg}
\end{eqnarray}
where $J=V > 0$ and $J_{xy} = -2 t_b < 0$ and $H = \mu - z/2 V$, $z$ is the
coordination number. 
%This model is antiferromagnetic
%along the $z$ direction and ferromagnetic along the perpendicular 
%direction.
% and its phase diagram for the parameters $H$ and $J_{xy}/J$
%is known for the square and triangular lattice by means of 
%various analytic and numerical means.\cite{batrouni,hebert,schmid}

% We wish to study the stability of the quantum
%solid with respect to the introduction of impurities. 
In order to illustrate the effects of the impurities on the stability
of the quantum solid and, vice versa, 
we first consider the square lattice because 
of its simplicity.
%Generalization of the approach to a triangular or hcp lattice
%is possible and it is the next step
%in order to make a closer connection to the experimental
%situation of supersolid helium.
Following the general spin-wave (SW) theory for an ordered square lattice
quantum antiferromagnet\cite{RMP}, we separate the 
ordered square lattice in two sub-lattices, A (or up, or occupied) 
and B (down, or empty) and
we consider boson operators $a^{\dagger}_i$ and $b^{\dagger}_i$ 
 which create  spin-deviations with respect to  the classical N\'eel 
ground state in sites of the corresponding sub-lattice. 
%Namely, we use the Holstein-Primakoff transformation,
%$S^+_i = a_i$, $S^z_i = {1/2} - a^{\dagger}_i a_i$ in the A sub-lattice
%and $S^+_i = b^{\dagger}_i$, $S^z_i =  b^{\dagger}_i b_i -{1/2}$,
%in the B sub-lattice. 
The Hamiltonian
(\ref{heisenberg}) is approximated by
keeping terms up to quadratic in spin-deviation operators;
using the Fourier transforms $a_{\bf k}$ and $b_{\bf k}$ of the operators 
$a_i$ and $b_i$ (as defined in Ref.~\cite{RMP}),
where ${\bf k}$ takes values from the Brillouin zone of the 
$\sqrt{2}\times \sqrt{2}$ sub-lattice and
introducing the Bogoliubov canonical transformation,
$a_{\bf k} = u_k \alpha_{\bf k} + v_k \beta^{\dagger}_{-\bf k}$,
$b_{\bf k} = u_k \beta_{\bf k} - v_k \alpha^{\dagger}_{-\bf k},$
where $\alpha^{\dagger}_{\bf k}$ and $\beta^{\dagger}_{\bf k}$
are boson creation operators, the Hamiltonian takes the form
\begin{eqnarray}
& & {\cal H}^B_{L}  =  E_0 + \sum_{{\bf k}} \Bigl (\omega^{\alpha}_{\bf k} 
\alpha^{\dagger}_{\bf k} a_{\bf k} + \omega^{\beta}_{\bf k}
 \beta^{\dagger}_{\bf k} 
\beta_{\bf k} \Bigr ),\label{bose-diagonal}
\end{eqnarray}
where $\omega^{\alpha,\beta}_{\bf k}  =   d J  \epsilon_{\bf k} \pm H$, 
$\epsilon_{\bf k}  =  \sqrt{ 1 - \lambda^2 \gamma^2_{\bf k}}$, 
$\gamma_{\bf k} = {1/2} (cos(k_x)+cos(k_y))$, $d=2$ and 
$\lambda= { {J_{xy}}/ J}$, and 
$u_{\bf k} = \sqrt{{1/2}({1/{\epsilon_{\bf k}}} + 1)}$,
$v_{\bf k} = - sgn(\gamma_{\bf k}) 
\sqrt{{1/2} ({1/{\epsilon_{\bf k}}}-1)}$. 

It follows from Eq.~\ref{bose-diagonal} that for
$\omega^{\alpha,\beta}_0  \ge 0$, i.e., for $H \ge d J \sqrt{1 - \lambda^2}$,
the N\'eel ordered ground state is unstable. Ferromagnetic (superfluid
in the Bose system) order 
develops in the $xy$ direction and the spins are canted
in order to acquire a component along the direction of the field. 
%When the magnetic field (chemical
%potential) is large, the magnetic system is paramagnetic and
%the bosonic system is in a band insulating
%state with filling factor $n=1$ (when $\mu > 0$) or 
%$n=0$ (when $\mu < 0$). When $|J_{xy}|/J < 1$ (i.e., $t_b/V<0.5$) 
%and in 2D for  $H /J < 2$ (i.e, $\mu <0$), it is
%an ordered solid with  checkerboard ($(\pi,\pi)$) ordering.
The phase diagram obtained for this model with this spin-wave approximation 
agrees reasonably well with that obtained by
other techniques.\cite{fisher-nelson}   
When the square lattice is half-filled there is a gap
$G=d J\sqrt{1-\lambda^2}$ for creating
a propagating pseudo-spin wave excitation, i.e.,  to promote
a boson atom to the interstitial band. These
interstitial quasiparticles move in a band which in the
limit of $J_{xy}<<J$ has a  bandwidth $W=dJ^2_{xy}/2J$.

\begin{figure}[ht] 
\begin{center}
\includegraphics[width=\figwidth]{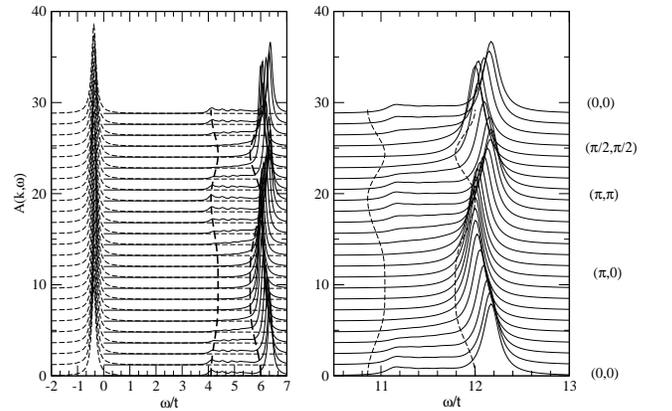}
 \caption{\label{fig1} 
The spectral functions $A({\bf k},\omega)$ for $J/t=3$
(left) and $J/t=6$ (right) along the 
Brillouin zone path
$(0.0) \to (\pi,0) \to (\pi,\pi) \to (0,0)$.}
\end{center}
\end{figure}
We wish to extend this approach to study the motion of a single impurity
inside the quantum solid.
As in Ref.~\cite{liu}, we consider as reference state the
N\'eel state with an impurity on the A (substitutional) or 
B (interstitial)  sub-lattices. We imagine 
that there exist operators $h^{\dagger}_i$ and $f^{\dagger}_i$ which
operate on the N\'eel state and replace respectively an up-spin or  
down-spin with an impurity.
For the single impurity case and by keeping only up 
to linear terms in spin-deviation operators the terms given by 
Eqs.~(\ref{impurity}) (\ref{interaction})  in terms of these operators
take the form
$\hat {\cal H}^*+\hat {\cal H}_{int} =  -  t \sum_{<ij>, i\in A} 
(a^{\dagger}_i f^{\dagger}_j h_i  + h.c.) +
 V^* d \sum_{ i \in B} f^{\dagger}_i f_i.$
The entire linearized Hamiltonian (\ref{model}), using the
the Bogoliubov transformation, takes the following form:
\begin{eqnarray}
&{\cal H}_L&= \sum_{\bf k} 
[\epsilon_1 h^{\dagger}_{\bf k} h_{\bf k}
+ \epsilon_2  f^{\dagger}_{\bf k} f_{\bf k}]
 +  \sum_{{\bf k}, {\bf q}} \Bigl [ g^{(1)}_{{\bf k} {\bf q}}
(f^{\dagger}_{\bf k-q} h_{\bf k}\alpha^{\dagger}_{\bf q} \nonumber \\
&+&h^{\dagger}_{\bf k }f_{\bf k -q}\alpha_{\bf q} )+g^{(1)}_{{\bf k}{\bf q}}
( f^{\dagger}_{\bf k}h_{\bf k -q} 
\beta_{\bf q}+h^{\dagger}_{\bf k -q }f_{\bf k}\beta^{\dagger}_{\bf q}) 
\Bigr ] + {\cal H}^B_{L}, \nonumber 
\end{eqnarray}
where 
%$\epsilon_1  =  V^* d \zeta$, $\epsilon_2 = V^* d (1-\zeta)$ and
$\epsilon_1  =  0$, $\epsilon_2 = V^* d $ and
$H^B_{L}$ is given by Eq. (\ref{bose-diagonal}).
The operators $f^{\dagger}_{\bf k}$ and
$h^{\dagger}_{\bf k}$  are the Fourier transforms of $f^{\dagger}_i$ and
$h^{\dagger}_i$ respectively and 
where $ g^{(1)}_{{\bf k}{\bf q}}  =  
 d \sqrt{2/N}  t \gamma_{\bf k-q} u_{\bf q}$, 
$g^{(2)}_{{\bf k}{\bf q}} = d \sqrt{2/N} t \gamma_{\bf k} v_{-\bf q}$.
%The red (blue) curves correspond to the substitutional
%(interstitial) impurities. The black and green solid lines are the
%dispersion of the peaks obtained from the variational calculation. }
\begin{figure}[ht] 
\begin{center}
\includegraphics[width=\figwidthb]{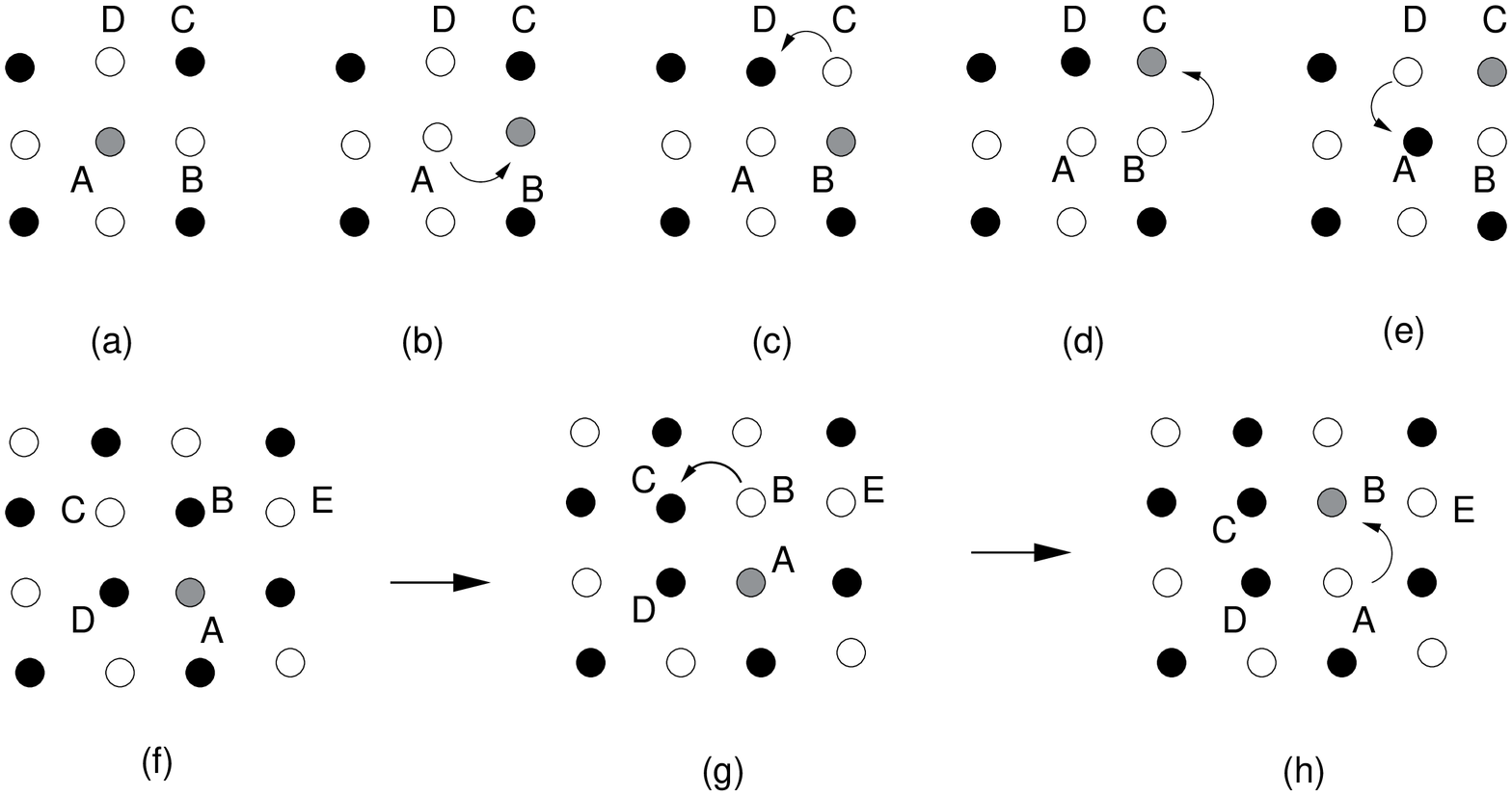}
 \caption{\label{fig2} 
The substitutional impurity moves by means 
of the $4^{th}$ order process (a)-(e). The
interstitial impurity moves via the $2^{nd}$ order
process (f)-(h). The impurity, bosonic atoms, and empty interstitial 
sites are denoted as gray solid, black solid, and open circles respectively.}
\end{center}
\end{figure}
\begin{figure}[ht] 
\begin{center}
\includegraphics[width=\figwidthb]{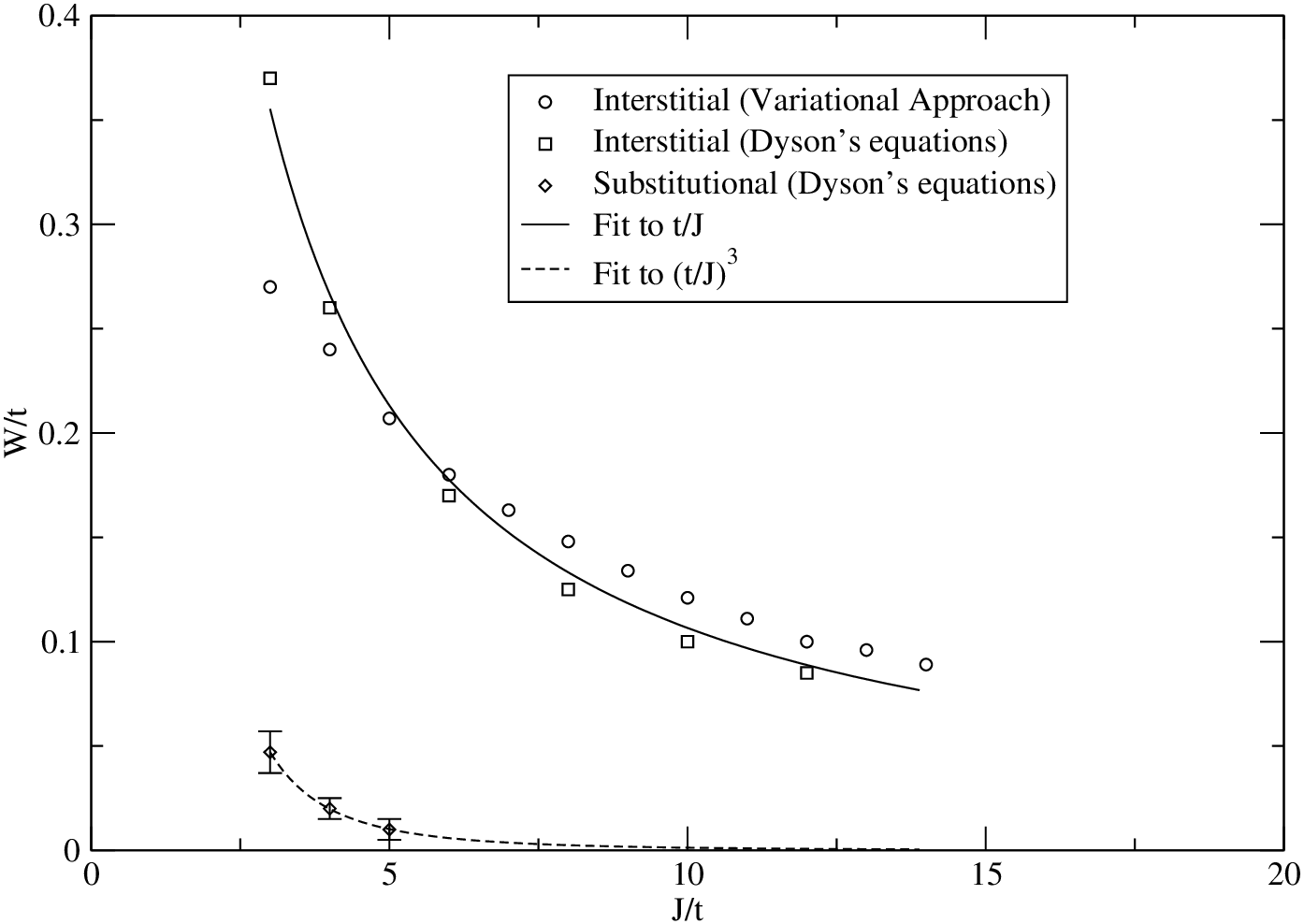}
 \caption{\label{fig3} 
The bandwidth of the substitutional and interstitial impurity bands
as a function of $J/t$.
}
\end{center}
\end{figure}
The two Dyson's equations in the non-crossing approximation 
are given as follows\cite{liu}:
\begin{eqnarray}
G_{\nu}({\bf k},\omega) =  {1\over
 {\omega - \epsilon_{\nu} 
 -  \sum_{\bf q} g^{(\nu)2}_{{\bf k} {\bf q}}
G_{\nu'}({\bf k-q}-\omega-\omega^{\mu}_{\bf q})}}, 
\label{dyson}
\end{eqnarray}
where the first equation is obtained for $\nu=1$, $\nu'=2$ and $\mu=\alpha$
and the second equation for $\nu=2$, $\nu'=1$ and $\mu=\beta$. Here 
$\omega_{\alpha,\beta}({\bf q})$ are given after 
Eq.~\ref{bose-diagonal}.
The Green's function $G_1$ ($G_2$) corresponds to the quasi-particles created
by $h^{\dagger}_{\bf k}$ ($f^{\dagger}_{\bf k}$). 
These equations can be solved iteratively as in Ref.~\cite{liu} 
starting from
$ G^{(0)}_{\nu}({\bf k},\omega) = 1/(\omega -\epsilon_{\nu} + i \eta)$.

In our calculations we took $V^*=V=J$, $J_{xy}=2 t$ (i.e., $t=t_b$), 
$H=0$ and $\epsilon=0.1$.
As discussed later NMR measurements\cite{nmr} indicate that for $^3He$ 
impurities in solid $^4He$ we should consider $J/t>>1$.
In Fig.~\ref{fig1} the spectral function $A({\bf k}=0, \omega)$
is presented for the case of $J/t=3$ (left) and $J/t=6$ (right). 
Notice that for  
$J/t=3$  the bandwidth $W_A$ of the substitutional impurity is small.
For $J/t=6$, $W_A$ is very small, i.e., $W_A/t \sim 10^{-3}$, and, hence, 
we only show the
spectral function of the interstitial impurity for this value of 
$J/t$ (right part of Fig.~\ref{fig1}). Notice that the spectral function
of the interstitial impurity  has 
two main peaks, a lower frequency peak with small spectral weight and 
a higher frequency one with most of the spectral weight. 
%The black and green solid lines are the dispersion of these
%two peaks as obtained from a variational calculation to be 
%discussed below. 

 In the regime of $J>>t$, the leading order
in $t/J$ which allows the substitutional impurity or a $^4He$ atom to move
is the fourth order process shown in Fig.~\ref{fig2}(a-e). 
Hence, the impurity moves in 
a band with a bandwidth of the order 
of $W_A/t = A (t/J)^3$ in an expansion of $t/J$ and $J_{xy}/J$ 
(which is also small  since $J_{xy}=2 t$). On the other hand, in our case where
$V^*=V=J$, the bandwidth of the interstitial impurity is
of the order of $W_B/t \sim t/J$ and it corresponds to the process
shown in Fig.~\ref{fig2}(f-h). States such as those of 
Fig.~\ref{fig2}(f) and Fig.~\ref{fig2}(h) 
are connected by second order degenerate perturbation theory processes
where the states shown in Fig.~\ref{fig2}(g) are included as intermediate
states.  
Namely, the interstitial impurity at site A takes advantage 
of a pair ``pseudo-spin''  flip near it, as in Fig.~\ref{fig2}(g), 
and hops  to site B as shown in Fig.~\ref{fig2}(h).

We also carried out a diagonalization in 
a space which includes the 17 states of the type shown in 
Fig.~\ref{fig2}(f-h) and their translations through the square lattice.
%In addition to the one of Fig.~\ref{fig2}(f) and 
%its translations, there are 12 states such as the state
%in Fig.~\ref{fig2}(g), where a 
%pair of pseudospins is flipped near the impurity. Furthermore, there are 
%4 states, such as the one in Fig.~\ref{fig2}(h), where the impurity
%is in a substitutional site and there is a pseudospin deviation in
%the interstitial sublattice nearest neighbor to the impurity.  
We found the dispersion 
%shown in Fig.~\ref{fig3}(bottom) 
%and it is also
indicated with the black and green lines in Fig.~\ref{fig1}. 
The bandwidth of the impurity as calculated both by solving 
the Dyson's Eqs.~\ref{dyson} and by the variational approach 
is shown in Fig.~\ref{fig3} as a function of $J/t$.  
Notice that the interstitial band (Fig.~\ref{fig1}) and bandwidth 
agree very well with those
obtained from the Dyson's equations. 
The solid line and dashed line are fits to $W_B/t = B (t/J)$
and $W_A/t = A (t/J)^3$ respectively with  $A \simeq B \simeq 1$.

Now, let us turn our discussion to the real case of $^3He$ impurities in 
solid $^4He$.
The form for bandwidth of the substitutional impurity,
i.e., $W_A \sim t (t/J)^3$
(when we take $t \sim t_b$ and $V^* \sim V$), 
is also valid on the triangular and hcp lattices; namely, when
the substitutional band is filled, in order for the
atoms to move, the same fourth order process, where the atoms
momentarily hop to interstitial positions, is necessary.
 In NMR\cite{nmr} studies  
tunneling rates were found to be of the order of 1 $MHz$. 
Using our calculated form for the bandwidth $W_A=A t (t/J)^3$, and taking
$t=1 K$ and $J=30 K$, we find $W_A \sim 4 \times 10^{-5} K$
(i.e., 1 $MHz$). Using our  form for the bandwidth for 
interstitial impurities $W_B=B t^2/J$ and the same value of $t$ and
$J$ we find $W_B \sim 40 mK$. Hence, 
the present theory
can reproduce (a) the NMR\cite{nmr} results, (b) that substitutional
quasiparticles might be localized\cite{kagan} is solid $^4He$, and (c) 
it suggests  that {\it interstitial} impuritons might move coherently
near the above temperature scale. However, as we discuss below,
interstitial impurities bind with defects.

For the case of interstitial $^4He$ atoms, using our result
for the bandwidth,
$W=dJ^2_{xy}/2J$ (and $J_{xy}
\sim 2 t$),  and the values of the parameters discussed in the previous
paragraph, we find $W\sim 200 mK$.
This is the temperature scale where
the possible super-solidity of $^4He$ has been observed\cite{chan}. 
In the case of real solid $^4He$ nearest neighbor empty
interstitial sites are much closer to one another than occupied sites
and, in addition, an interstitial atom has higher energy
than a substitutional atom and therefore faces a lower
potential barrier to tunnel to another interstitial site.
These two factors increase the tunneling frequency by several orders
of magnitude relative to the observed frequency\cite{nmr} for
substitutional sites. 
%In recent neutron scattering studies 
%from the bcc crystal\cite{pelleg}, an 
%optic-like mode has been observed at about $10 K$ with 
%a dispersion-bandwidth of the order of $1 K$.  
However, a finite energy  is needed in order to promote atoms 
to the interstitial band\cite{dai} and, in
addition, vacancies and interstitials have a tendency to phase 
separate in an equilibrated crystal\cite{vacancy}.
On the other hand, inhomogeneities provided by  $^3He$ impurities 
during the $^4He$ solid nucleation process or by a relatively
fast inhomogeneous cooling process can create such interstitial atoms as 
discussed next.

%Now, let us discuss what happens to an impurity in solid $^4He$.
\begin{figure}
%\vskip 0.1 in
\begin{center}
\includegraphics[width=\figwidth]{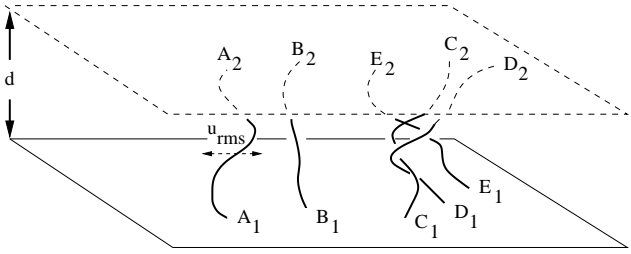}
 \caption{
Edge-dislocations or disclication lines form as the $^4He$ solid grows on the 
cell surface (bottom plane). The top (dashed) plane
indicates the solid/liquid interface. 
We imagine that the solid grows layer by layer. Because the
interface is two-dimensional, dislocations and disclinations might be expected
to appear during a relatively
fast cooling process at any given layer. In addition, the impurities
present in the liquid bind very quickly to such point defects.
The point defects of the next layer 
more or less line up with those of the previous layer (otherwise
it is energetically very costly). Therefore, these defects
become lines of defects as shown in the above figure. 
When the average distance between such defects  is smaller than 
the length $u_{rms}$ which characterizes the wandering of each 
defect line,  the defect-lines can entangle (such as the lines 
$C_1C_2$, $D_1D_2$ and $E_1E_2$)
and can produce a topological glass phase.}
\label{fig4}
\end{center}
%\vskip 0.1 in
\end{figure}
As already discussed, the model studied in this paper is very similar
to the $t-J$ model used to understand quantum antiferromagnets,
where the impurity degrees of freedom in our case map to 
the holes in the $t-J$ model and the hopping of the bosonic atoms maps
to the spin degrees of freedom. In the case of the
$t-J$ model on the square lattice, when a finite density of holes 
is introduced  the ground state may have 
stripe-order\cite{white} where the stripes
are hole-rich and the domains between stripes are 
antiferromagnetically ordered (quantum solid domains) with a $\pi$-phase
shift between domains. 
As we will argue below, a similar phenomenon should be expected to occur 
in the case of an hcp lattice of bosons with impurities.
Namely, edge-dislocations or disclination lines such as those shown in 
Fig.~\ref{fig4} might be created during the process of
solid $^4He$ growth  from the liquid phase. 

First, it is known that the inhomogeneous 
nucleation of solid $^4He$ starts on the cell walls\cite{balibar}.
We will assume that solid $^4He$ grows layer-by-layer 
on the surface of the cell for the following reasons:
(a) The geometry of the cells used in the experiments\cite{chan,reppy2} 
is characterized by a large surface with small distances between
large confining walls. (b) The cooling of the solid is caused through
cooling of these large surfaces, therefore, atoms near the surface will
get cold first.
(c) Helium atoms at temperature $T\sim 1 K$ 
get adsorbed on the surface of most substrates because of a
relatively strong dipolar interaction with the surface\cite{cole}. 
In studies of 
$^4He$ on graphite\cite{marlon,crowell} for example and in all known substrates
the first layer of adsorbed $^4He$ is solid. The second layer of
$^4He$ on the graphite surface is also solid at layer completion 
density\cite{marlon2,crowell}. After the first or a few layers are deposited
on the surface, the adsorbed solid $^4He$ layer is {\it compressed} 
relative to the liquid in contact and, therefore, the coated surface provides
a stronger attractive potential for an atom as compared to that provided by its
surrounding atoms in the liquid; thus, the atoms tend to get adsorbed
and form a new solid layer.
Nucleation of solid $^4He$ had been a puzzle in the past 
just because of the large surface energy cost  required in
homogeneous nucleation\cite{balibar}.
 In a layer-by-layer growth this is not an issue, because when 
the next layer is deposited, there is no
additional surface area introduced, namely, just the same-size 
interface advances.
 
It is believed\cite{carmi} that  the solid/liquid interface
in the presence of $^3He$ impurities becomes rough
due to binding of such impurities to the interface. Even an extremely
low impurity concentration, namely as low as 1 ppm, has a significant effect
in this process\cite{carmi}.
In this paper, it is proposed that the $^3He$ impurities bind to 
defects\cite{bishop}, namely to dislocations or disclinations,
which are expected to form on the two-dimensional solid/liquid 
interface\cite{nelson}. 
There are various theories of two dimensional 
melting\cite{kleinert}, the most popular
of which is the theory of Halperin, Nelson and Young\cite{nelson} which
was inspired by the Kosterlitz-Thouless(KT) theory of vortices in
superfluid films. According to this theory, the melting proceeds 
via two continuous KT transitions, the lowest temperature one caused by the
unbinding of dislocations and a higher temperature transition caused 
by the unbinding of disclinations.
Both types of defects are point-like singularities; the former corresponds
to the singular behavior of the atomic lattice displacement 
field  and the latter is associated with the singular behavior of the angle 
field $\theta(\vec r)$  which characterizes
the fluctuations in the bond orientational order. 
Our following analysis applies equally to both type of defects and, therefore,
we will use the term {\it defect} in general.
As the solid/liquid interface is cooled down at a relatively fast 
rate, locally  small-size 2D clusters of  atoms at the solid/liquid 
interface achieve local crystalline order rather quickly;
however, the time scale to achieve a global equilibrium
is much longer because these clusters may have different orientations.
In order to describe the state of this globally disordered
but locally ordered 2D system, we need to imagine the presence of 
dislocations
(or disclinations). Generally, as a 2D disordered solid is cooled down to
low temperature, pairs 
of such thermally created excitations having opposite
topological charge move very slowly towards each other and are 
annihilated (or bind). However, the $^3He$ impurities which are present
in the liquid find it energetically favorable to bind to these defects
and the combined impurity/defect system becomes stable.
Furthermore, it becomes energetically favorable
for these lattice defects of the next deposited
solid layer to follow the positions of the defects on 
the previous layer, thus, defect-lines are formed as the 
interface advances; namely,  lines of singularities grow
perpendicular to the interface (see Fig.~\ref{fig4}). 

The root-mean-square projection $u_{rms}$ of the end-to-end vector of 
a defect-line
is given as\cite{nelson-seung}
$u_{rms}=\sqrt{<u^2>}=(2 \pi k_B T d/\epsilon)^{1/2}$ where $d$ 
is the thickness of the
solid film and $\epsilon$ is the line tension along the 
defect-line. 
When $u_{rms}$ is a few times greater than
the average distance between such defects, neighboring 
defect-lines will entangle. 
Taking $T\sim 1 K$ and $\epsilon \sim 10 K/\AA$ and $d=0.3 mm$ we find
that $u_{rms}\sim 10^3 \AA$ which is greater than the average distance
between impurities and, thus, entanglement seems likely.
When the solid is cooled to low temperature
defects with opposite topological charge cannot be mutually
annihilated because the entangled defects cannot move. The
entanglement is topologically protected and this could create a topological
glass state of solid $^4He$. This is analogous to the vortex-glass state
proposed for superconductors\cite{nelson2,fisher}.

%a substitutional site by promoting a $^4He$ atom to an
%interstitial position. The reason is that
%the $^3He$ atomic mass is smaller than the $^4He$ mass,
%therefore, the configuration in which the $^3He$ atoms is
%``squeezed'' in the interstitial space has significantly higher
%energy than the one where the $^4He$ is in the interstitial space.
%For a similar reason even a substitutional $^3He$ atom has a tendency to 
%promote a $^4He$ atom to an interstitial space as illustrated in 
%Fig.~\ref{fig4}.
%A substitutional $^3He$ impurity in solid $^4He$ prefers a larger
%room for zero-point motion than the solid $^4He$ lattice
%spacing because of its lighter mass. As a result,
%it causes a local expansion which means that the atoms labeled as 1-6
%in the Fig.~\ref{fig4} will spend some time as interstitial $^4He$ atoms.
%This implies that $^3He$ impurities are expected to promote
%$^4He$ atoms from the substitutional band which is full to an
%empty interstitial band with a relatively large bandwidth.

As discussed earlier $^3He$ impurities, even at
concentrations as low as 1 ppm, influence very significantly 
the solid $^4He$ nucleation process\cite{carmi}. In particular 
even such very small amount of impurities can change the
roughening transition temperature\cite{carmi} by $20 \%$. 
This has been interpreted as the result of $^3He$ impurity adsorption
at the liquid to solid interface. As discussed such impurities
stabilize the defects and as the interface advances defect
lines form which may entangle and this can
influence significantly the roughening transition. 
Therefore, it is possible that the $^3He$ impurities, through the
creation of such defects, promote $^4He$ atoms to the interstitial band
which can move from domain to domain and by means of the entangled 
defects.
This creates a metastable disordered supersolid in which the carriers 
flow through a topological glass of entangled defect-lines.
This scenario  might explain the observed hysteresis\cite{reppy} by
annealing of the solid helium sample. 
Most recently  Rittner 
and Reppy\cite{reppy2}  have observed very high superfluid response
in their torsional oscillator experiments of solid $^4He$ samples grown 
from the liquid phase by very rapid cooling. Furthermore, these
samples have high surface-to-volume-ratio and high concentration of
such defects is expected to occur under such conditions.
Namely, under rapid cooling such defects
are expected to occur in the 2D solid/liquid interface, because,
the atoms order locally to form a microscopic-size solid very quickly, 
but the time scale for 
the annihilation of pair of defects of opposite topological
charge is very long, namely much longer than the time required 
to establish local (i.e., at a microscopic scale) equilibrium.
Once these defect-lines are formed, they wander around as
shown in Fig.~\ref{fig4} and they may entangle around each other
several times because of their high density. The process of 
annihilation of these defects is then very difficult because of their
entanglement  and this may
lead to an extremely long-lived topological glass phase.

Experimentally, while the critical temperature $T_c$ increases with 
increasing $^3He$ impurity concentration $x_3$, for the superfluid 
response $\rho_s$  there is an optimum $x_3$
 above which $\rho_s$ decreases with increasing $x_3$.
This behavior is consistent with the theory presented here.
The degree of inhomogeneity is proportional to the density of defects
which is also proportional to the impurity concentration $x_3$.
The carrier density, namely the density of interstitial atoms,
increases with $x_3$, therefore, $T_c$ is expected to increase with $x_3$.
Since $T_c$ is more or less a locally determined quantity, the 
effect of global disorder on $T_c$ is much less important than the fact 
that the number of carriers rise with disorder.
The superfluid response, however, is expected to depend strongly on 
phase fluctuations of the superfluid order parameter, therefore, disorder
is expected to have harmful effects to the long-range coherence.
In summary, it is reasonable to expect that
at high concentration of such $^3He$ impurities, the 
disorder which is caused by the lattice defects should significantly 
harm the  long-range coherence, however, $T_c$ may continue to rise
because, while the long-range disorder does not significantly influence
the value of $T_c$, $T_c$ increases due to the increase of the 
number of carriers.
%because the state of solid $^4He$ with interstitial 
%$^3He$ impurities should exist only as a metastable phase.

%If indeed $^3He$ impurities
%drive the superfluid behavior in solid $^4He$, 
%the condensate fraction, naively, should 
%be proportional to the $^3He$ impurity concentration. Therefore,
%the large difference between condensate
%fraction and the observed superfluid fraction needs to be explained.
%A large difference between these
%cannot be excluded because we have examples, such as liquid $^4He$
%near solidification and at low temperature, where the condensate 
%fraction is small and the superfluid fraction is almost $100\%$.
%However, a simple explanation from the present work cannot be provided 
%and further experimental
%and theoretical work is needed.
%Some of the proposed ideas can be tested using path integral Monte
%Carlo simulation\cite{wierschem}.

I would like to thank 
%P. Schlottmann for discussions and 
K. Wierschem for proof-reading the manuscript and M. Boninsegni for
useful discussions.
This work was supported by NASA grant NAG-2867.

\end{document}